\documentclass[11pt,twoside]{article}
\usepackage{macro-fsut-eng}
\usepackage{graphicx}

\usepackage[T1]{fontenc} 

\usepackage{latexsym}
\usepackage{verbatim}
\usepackage{hyperref}

\begin{document}

\title{A Rastall Scalar-Tensor theory}

\author{Thiago Caram\^es, J\'ulio C. Fabris, Oliver F. Piattella, Vladimir Strokov\footnote{On leave from the Lebedev Physical Institute (Moscow, Russia)}}
\affil{Departamento de F\'{\i}sica - UFES, Vit\'oria, ES, Brazil}

\author{Mahamadou H. Daouda}
\affil{D\'epartament de Physique - Universit\'e de Niamey, Niamey, Niger}

\author{Adriano M.  Oliveira}
\affil{IFES, Guarapari, ES, Brazil}

\date{\today}

\begin{abstract}
We formulate a theory combining the principles of a scalar-tensor gravity and the Rastall proposal of a violation of the usual conservation laws.
In the resulting Brans-Dicke-Rastall (BDR) theory the only exact, static, spherically symmetric solution is a Robinson-Bertotti type solution besides the trivial Schwarzschild one.  The PPN constraints can be completely satisfied for some values of the free parameters.The cosmological solutions display, among others, a decelerate-accelerate transition in the matter dominated phase.
\end{abstract}

\section{Introduction}

The Brans-Dicke theory, the paradigma of scalar-tensor theories, is considered as an important alternative to the theory of General Relativity (GR) [Brans\&Dicke, 1962]. In this theory
the gravitational coupling $G$ is considered as a dynamical quantity represented by the field $\phi$ which is introduced in the gravitational action through a kinetic term and a non-minimal coupling with the usual Ricci scalar.
A new parameter $\omega$ quantifies the interaction of the scalar field and the gravitational term, such that as $\omega \rightarrow \infty$ the General Relativity
theory is recovered. Recent estimates using the PLANCK data indicates a value $\omega \sim 1000$ [Avilez\&Skordis, 2013]. In spite of those observational constraints, small -- or even negative -- values of the parameter $\omega$ may be very interesting. They arise, for example, in the string theories in their low-energy limit [Lidsey  et al, 2000]. When negative values of $\omega$ are allowed, primordial singularity-free solutions emerge naturally from
the Brans-Dicke theory [Gurevich et al, 1973]. Late time accelerated solution can also be achieved, but at the price of a negative gravitational coupling [Batista et al, 2001].
\par
Some generalisations of GR evoke the gravitational anomaly effect, viz. Rastall's theory. [Rastall, 1972; 1976] These generalisations touch one of the cornerstones of gravity theories: the conservation laws encoded in the null divergence of the energy-momentum tensor. Since the concept of energy in GR is an object of discussion, the possibility that the energy-momentum tensor has a non-zero divergence should be considered in some situations.
The idea of violation of the conventional conservation laws in the context of Brans-Dicke theory has been considered by Smalley [Smalley, 1974]. In this approach, the Klein-Gordon type equation for the scalar field was kept as in the Brans-Dicke theory while the Einstein equations were changed accordingly. Here, we would like to revisit this proposal following a different path: we try to write down the field equations in such a way that the Brans-Dicke, GR as well as the ordinary the Rastall's theory are recovered: we keep the violation of the energy-momentum tensor in a spirit very close to the original formulation of the Rastall's theory, and the Klein-Gordon type equation as well as the Einstein's equations are modified accordingly.
\par
We investigate the resulting theory in two situations: spherically symmetrical and cosmological configuration. In the former case, we obtain that the only non-trivial solution is represented by the Robinson-Bertotti metric [Bertotti, 1959; Robinson, 1959] (its interpretation, however, differs from the conventional one). A solution that represents a star-like configuration is the ``trivial'' Schwarzschild one. At cosmological level, we show that accelerated solutions are possible in the dust phase of the cosmic evolution without introducing dark energy. We display a particular case where a decelerated/accelerated transition in the recent universe is achieved with a positive effective gravitational coupling.

\section{The theory}\label{theory}

The main idea of Rastall's theory [Rastall, 1972] is the assumption that in curved space-time the usual conservation laws used in GR are violated.
Hence, there must be a connection between the divergence of the energy-momentum tensor and the curvature of the space-time.
According to this program, the divergence of the energy-momentum tensor may be written as
\begin{eqnarray}
\label{r1a}
{T^{\mu\nu}}_{;\mu} = \frac{1 - \lambda}{16\pi G}R^{,\nu}.
\end{eqnarray}
In equation (\ref{r1a}) $\lambda$ is a free parameter codifying the deviation from the conservation. When $\lambda = 1$ the traditional conservation laws are recovered. Equation (\ref{r1a}) is a phenomenological way to implement the gravitational anomaly due to quantum effects.
\par
In the context of the Brans-Dicke theory, we can make the identification $
G \propto \frac{1}{\phi}$.
Hence,
\begin{eqnarray}
\label{h1}
{T^{\mu\nu}}_{;\mu} = \frac{(1 - \lambda)\phi}{16\pi}R^{,\nu}.
\end{eqnarray}
Let us generalize Rastall's version of the field equations to the Brans-Dicke case. Following the original Rastall's formulation in the context of GR, we write,
\begin{eqnarray}
\label{h2}
R_{\mu\nu} - \frac{\lambda}{2}g_{\mu\nu}R = \frac{8\pi}{\phi}T_{\mu\nu}
+ \frac{\omega}{\phi^2}\biggr\{\phi_{;\mu}\phi_{;\nu} - \frac{1}{2}g_{\mu\nu}\phi_{;\rho}\phi^{;\rho}\biggl\}
+ \frac{1}{\phi}\biggr\{\phi_{;\mu;\nu} - g_{\mu\nu}\Box\phi\biggl\}.
\end{eqnarray}
Combining the hypothesis (\ref{h1}) and (\ref{h2}), and using the Bianchi's identities, we obtain that the scalar field $\phi$ must obey the equation:
\begin{eqnarray}
\label{h3}
\Box\phi = \frac{8\pi\lambda}{3\lambda - 2(1 - 2\lambda)\omega}T - \frac{\omega(1 - \lambda)}{3\lambda - 2(1 - 2\lambda)\omega}\frac{\phi^{;\rho}\phi_{;\rho}}{\phi}.
\end{eqnarray}
Equations (\ref{h1},\ref{h2},\ref{h3}) form our complete system in this new formulation.
When $\lambda = 1$, the usual Brans-Dicke theory is recovered.
\par
The effective gravitational coupling today reads:
\begin{eqnarray}
\label{efetivo}
G = \frac{2[2\lambda + (3\lambda - 2)\omega)]}{3\lambda - 2(1 - 2\lambda)\omega}\frac{1}{\phi}.
\end{eqnarray}
When $\lambda = 1$ we obtain the corresponding expression for the Brans-Dicke theory.

\section{Spherically symmetric static vacuum solutions}\label{spherical}

The classical tests of theory of gravity are based on the motion of test particles in the geometry of a spherically symmetric object like a star or a planet.
Hence, to verify the viability of the theory proposed, it is crucial to look for a spherically symmetric solution. As a first step, the (exterior) solution representing
the space-time of a star-like object is considered.
\par
In the vacuum case, the equations reduce to
\begin{eqnarray}
R^{,\nu} &=& 0,\\
R_{\mu\nu} - \frac{1}{2}g_{\mu\nu}R &=&
\frac{\omega}{\phi^2}\biggr\{\phi_{;\mu}\phi_{;\nu} + \frac{\lambda}{2(1 - 2\lambda)}g_{\mu\nu}\phi_{;\rho}\phi^{;\rho}\biggl\}\nonumber\\ &+& \frac{1}{\phi}\biggr\{\phi_{;\mu;\nu} +
\frac{(1 + \lambda)}{2(1 - 2\lambda)}g_{\mu\nu}\Box\phi\biggl\},\\
\label{v-e3}
\Box\phi &=& - \frac{\omega(1 - \lambda)}{3\lambda - 2(1 - 2\lambda)\omega}\frac{\phi^{;\rho}\phi_{;\rho}}{\phi}.
\end{eqnarray}
The first of these equations leads to $R = R_0 = \mbox{constant}$.
Hence, in vacuum the Ricci scalar is necessarily constant. The case $R_0 = 0$ corresponds to the Schwarzschild solution of GR.
\par
Let us consider a metric in the form:
\begin{eqnarray}
ds^2 = e^{2\gamma}dt^2 - e^{2\alpha}dr^2 - e^{2\beta}(d\theta^2 + \sin^2\theta d\phi^2).
\end{eqnarray}
The functions $\alpha$, $\beta$ and $\gamma$ depend on the radial coordinate $r$ only.
First we find that the constant $R_0$ is given by:
\begin{equation}
\label{curva}
R_0 = \omega\biggr\{\frac{3 + 2\omega}{3\lambda - 2(1 - 2\lambda)\omega}\biggl\}\frac{\phi_{;\rho}\phi^{;\rho}}{\phi^2}.
\end{equation}
Now, the D'Alambertian reads:
\begin{equation}
\label{box1}
\Box\phi = \frac{\left(\sqrt{-g}g^{\mu\nu}\phi_{,\nu}\right)_{,\mu}}{\sqrt{-g}} = - e^{-2\alpha}[\phi'' + (\gamma' + 2\beta' - \alpha')\phi']\,,
\end{equation} 
where $g\equiv\det{g_{\mu\nu}}$.
\par
Let us choose the radial coordinate such that
$\alpha = \gamma + 2\beta$.
Writting the Einsteinian equations as
\begin{eqnarray}
\label{eeq}
R_{\mu\nu} = \frac{\omega}{\phi^2}\biggr\{\phi_{;\mu}\phi_{;\nu} + \frac{\lambda - 1}{2(1 - 2\lambda)}g_{\mu\nu}\phi_{;\rho}\phi^{;\rho}\biggl\}
+ \frac{1}{\phi}\biggr\{\phi_{;\mu;\nu} + \frac{\lambda - 2}{2(1 - 2\lambda)}g_{\mu\nu}\Box\phi\biggl\}\,,
\end{eqnarray}
we obtain, in the extended form:
\begin{eqnarray}
\label{fe-1}
\gamma'' + \frac{\phi'}{\phi}\gamma' = - \omega\frac{(\lambda - 1)}{2(1 - 2\lambda)}\biggr(\frac{\phi'}{\phi}\biggl)^2 - \frac{\lambda - 2}{2(1 - 2\lambda)}\frac{\phi''}{\phi}, \\
\label{fe-2}
\gamma'' + 2\beta'' - 2\beta'(\beta' + 2\gamma') - \frac{\phi'}{\phi} (\gamma' + 2\beta') = - \omega\frac{1 - 3\lambda}{2(1 - 2\lambda)}\biggr(\frac{\phi'}{\phi}\biggl)^2 + \frac{3\lambda}{2(1 - 2\lambda)}\frac{\phi''}{\phi},\\
\label{fe-4}
\beta'' + \beta'\frac{\phi'}{\phi} - e^{2(\gamma + \beta)} = -\omega\frac{\lambda - 1}{2(1 - 2\lambda)}\biggr(\frac{\phi'}{\phi}\biggl)^2 - \frac{\lambda - 2}{2(1 - 2\lambda)}\frac{\phi''}{\phi}.
\end{eqnarray}

Equations (\ref{v-e3},\ref{curva}) lead to two supplementary equations:
\begin{eqnarray}
\label{r0}
R_0 &=& - \omega\biggr\{\frac{3 + 2\omega}{3\lambda - 2(1 - 2\lambda)\omega}\biggl\}e^{-2\alpha}\biggr(\frac{\phi'}{\phi}\biggl)^2,\\
\phi'' &=& - \omega\frac{1 - \lambda}{3\lambda - 2(1 - 2\lambda)\omega}\frac{\phi'^2}{\phi}.
\end{eqnarray}
 
The only self-consistent solution for the above equations are:
\begin{eqnarray}
\label{alpha_beta_gamma}
\alpha &=& \alpha_0 - \ln{(r/r_0)},\\
\gamma &=& \gamma_0 - \ln{(r/r_0)},\\
\beta &=& \beta_0 = \frac{1}{2}(\alpha_0 - \gamma_0),\\
\phi &=& \phi_{0} (r/r_0)^\frac{1}{1 - A}, \quad A = - \omega\frac{1 - \lambda}{3\lambda - 2(1 - 2\lambda)\omega}.
\end{eqnarray}
Hence, the metric is
\begin{equation}
ds^2 = e^{2\alpha_0}\frac{dt^2}{(r/r_0)^2} - e^{2\gamma_0}\frac{dr^2}{(r/r_0)^2} - e^{\alpha_{0}-\gamma_{0}}d\Omega^2.
\end{equation}
If the scale~$r_0$ is chosen such that $r_{0}^{2}=e^{\alpha_{0}-3\gamma_{0}}$, making redefinitions $t\rightarrow e^{-(\gamma_{0}+\alpha_{0})/2}$, $s\rightarrow e^{-(\gamma_{0}-\alpha_{0})/2}$, and $r\rightarrow rr_{0}$, we arrive at:
\begin{equation}
\label{sol-1}
ds^2 = \frac{1}{r^{2}}\left(dt^2 - dr^2 - r^{2}d\Omega^2\right)\,,
\end{equation}
which is the so-called Robinson-Bertotti solution that is obtained, in the context of GR, by considering an electromagnetic field. Hence, no black hole solution is possible. This solution appears as the only non-trivial (non-Schwarzschild solution) vacuum solution.

A PPN analysis reveal that the classical tests of a gravitation theory are equally satisfied as in Genera Relativity if $\lambda = 0$ [Caram\^es et al, 2014]. This makes this theory quite competitive.

\section{Cosmology}\label{cosmology}

Let us consider an isotropic and homogeneous space-time described by the flat Friedmann-Lema\^{\i}tre-Robertson-Walker (FLRW) metric,
\begin{equation}
ds^2 = dt^2 - a(t)^2(dx^2 + dy^2 + dz^2),
\end{equation}
and an equation of state of the type $p = \alpha\rho$, with $\alpha =$ constant.
In this case the equations of motion read:
\begin{eqnarray}
\label{em1}
\dot\rho + 3\frac{\dot a}{a}(1 + \alpha)\rho = - \frac{3(1 - \lambda)}{8\pi}\phi\biggr[\frac{\stackrel{...}{a}}{a} + \frac{\dot a}{a}\frac{\ddot a}{a} - 2\biggr(\frac{\dot a}{a}\biggl)^3\biggl],\\
\label{em2}
3\biggr(\frac{\dot a}{a}\biggl)^2 = \frac{8\pi \rho}{\phi}\biggr\{\frac{1 - 3\lambda}{2(1 - 2\lambda)} + \frac{3(1 - \lambda)}{2(1 - 2\lambda)}\alpha\biggl\}
+ \omega\biggr[\frac{2 - 3\lambda}{2(1 - 2\lambda)}\biggl]\biggr(\frac{\dot\phi}{\phi}\biggl)^2\nonumber\\
+ \biggr[\frac{3(1 - \lambda)}{2(1 - 2\lambda)}\frac{\ddot\phi}{\phi} + \frac{3(1 + \lambda)}{2(1 - 2\lambda)}\frac{\dot a}{a}\frac{\dot \phi}{\phi}\biggl],\\
\label{em3}
2\frac{\ddot a}{a} + \biggr(\frac{\dot a}{a}\biggl)^2 = - \frac{8\pi}{\phi}\biggr\{\frac{1 - \lambda - (1 + \lambda)\alpha}{2(1 - 2\lambda)}\biggl\}\rho + \omega\frac{\lambda}{2(1 - 2\lambda)}\biggr(\frac{\dot\phi}{\phi}\biggl)^2 \nonumber\\
+ \frac{1 + \lambda}{2(1 - 2\lambda)}\frac{\ddot \phi}{\phi} + \frac{5 - \lambda}{2(1 - 2\lambda)}\frac{\dot a}{a}\frac{\dot\phi}{\phi},\\
\label{em4}
\frac{\ddot\phi}{\phi} + 3\frac{\dot a}{a}\frac{\dot\phi}{\phi} =  \frac{8\pi\lambda}{3\lambda - 2(1 - 2\lambda)\omega} (1 - 3\alpha)\frac{\rho}{\phi}
- \omega\frac{1 - \lambda}{3\lambda - 2(1 - 2\lambda)\omega}\biggr(\frac{\dot\phi}{\phi}\biggl)^2.
\end{eqnarray}
\par
Equations (\ref{em1})-(\ref{em4}) form a rich and complex system. In order to get a hint on which kind of solutions they predict, we consider power-law solutions, in the first place. The power-law solutions constitute a very restrictive case, but they can indicate the kind of cosmological solution we can expect from the BDR theory. Hence, suppose the solutions have the form $a = a_0t^s, \phi = \phi_0t^p, \rho = \rho_0t^q$,
where $a_0$, $\phi_0$, $\rho_0$, $s$, $p$ and $q$ are constants.
This system admits eight pairs of roots for $(s,p)$. For the dust case, $\alpha = 0$, one of the pairs corresponds to the Minkowski case, $p = s = 0$. Another one is $s = p = 1/2$. A third root implies a curious configuration with $s = 0$ and $p = 2$, that is, a static universe, with a varying gravitational coupling. Among the other five pairs, two incorporate an accelerated regime of the expansion while remaining three describe a decelerating universe. The overall situation is described in reference [Caram\^es et al, 2014], where the meaning of these different results are analysed.
\par
The field equations depends not only on the values of $\lambda$ and $\omega$, but also on the value of the initial conditions. We look for an example of a deceleration/acceleration transition during the matter dominated phase ($\alpha = 0$). Figure 1 shows the behaviour of the Hubble function $H = \frac{\dot a}{a}$ and deceleration parameter $q = - 1 - \frac{\dot H}{H^2}$, for $\omega = 1$ and $\lambda = - 1$, undergoing this transition. Note that the effective $G > 0$ stays positive~(see~(\ref{efetivo})).

\begin{center}
\begin{figure}[!t]\label{fig1}
\begin{minipage}[t]{0.2\linewidth}
\includegraphics[width=\linewidth]{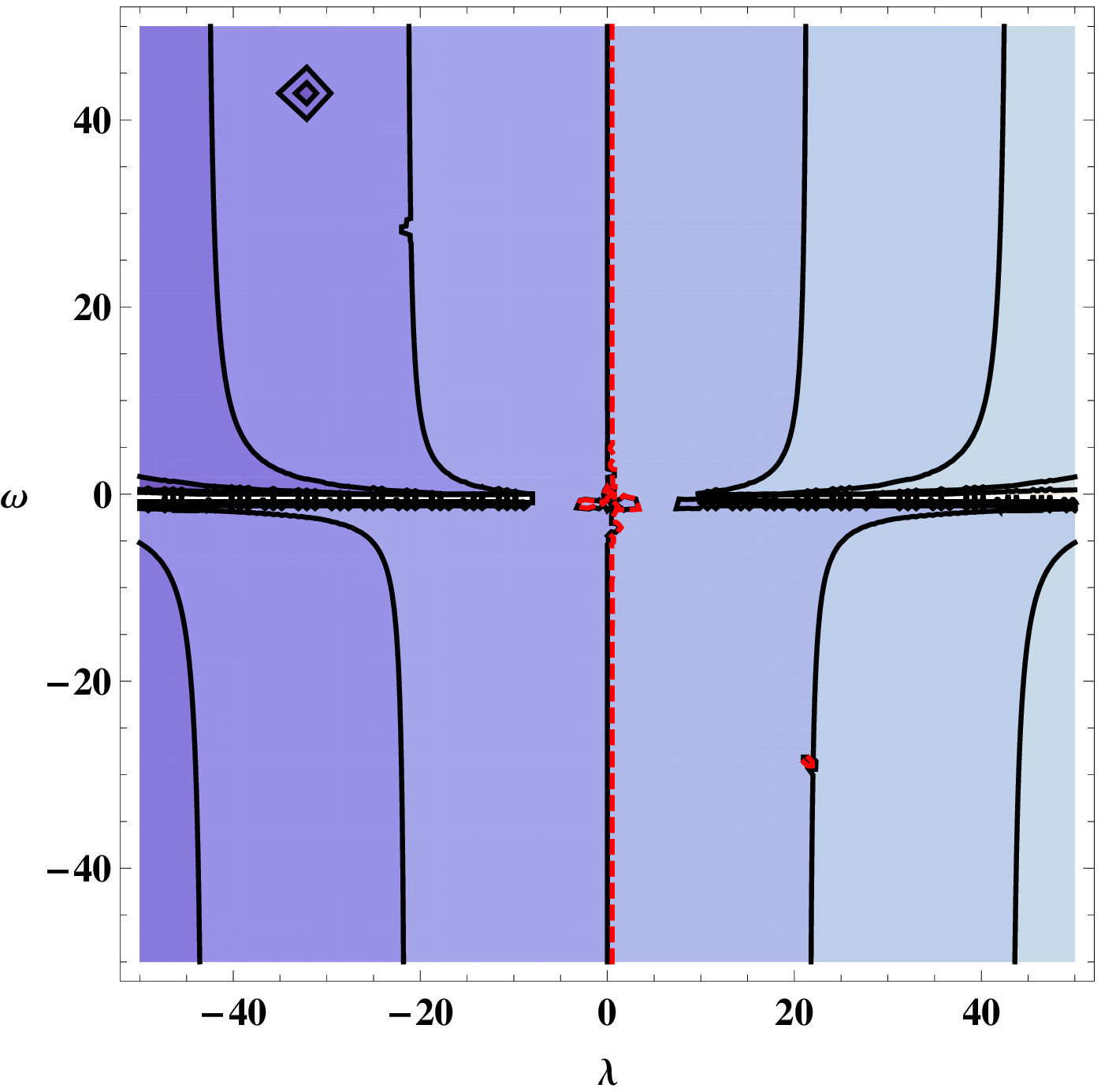}
\end{minipage} \hfill
\begin{minipage}[t]{0.2\linewidth}
\includegraphics[width=\linewidth]{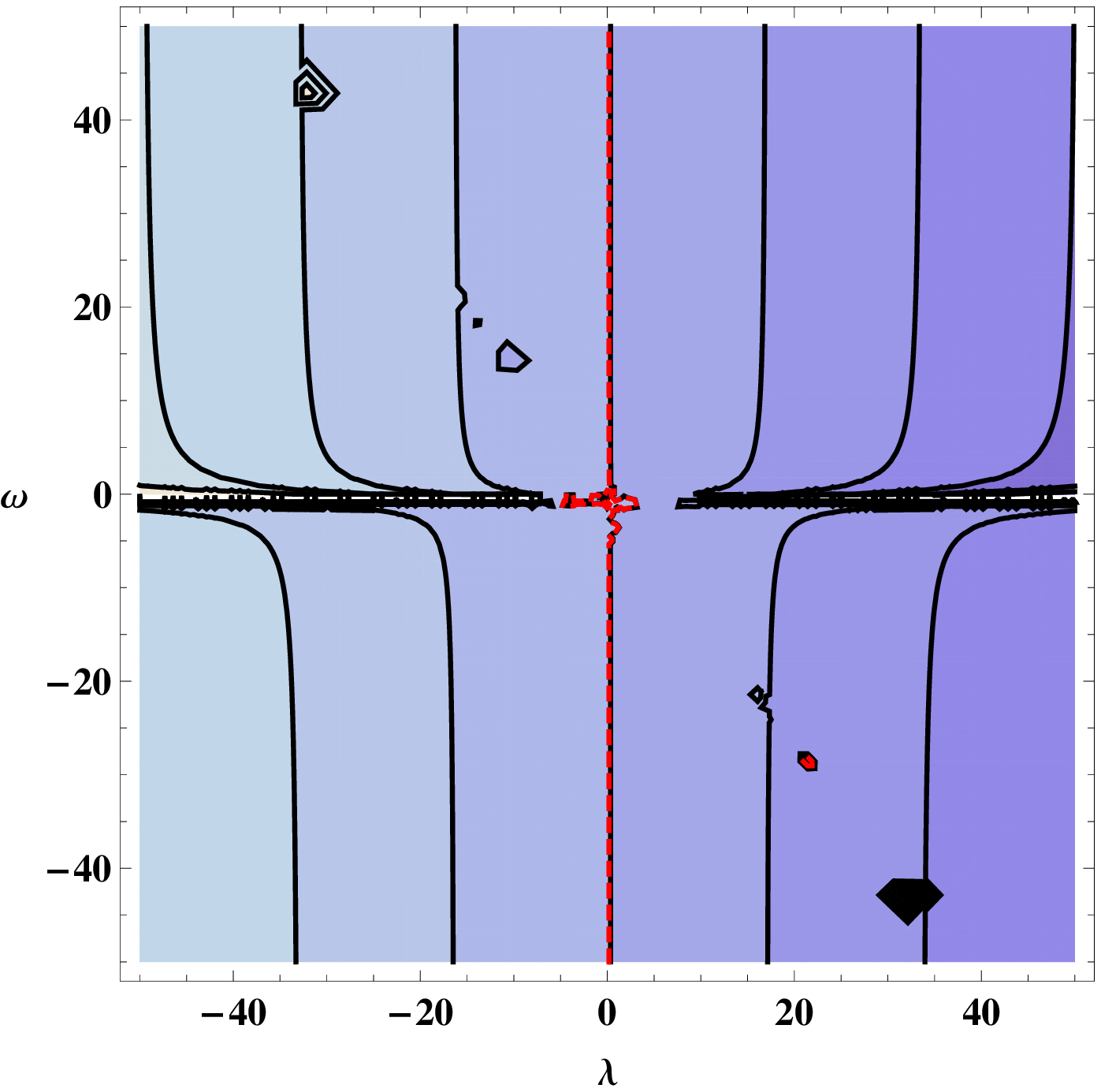}
\end{minipage} \hfill
\begin{minipage}[t]{0.2\linewidth}
\includegraphics[width=\linewidth]{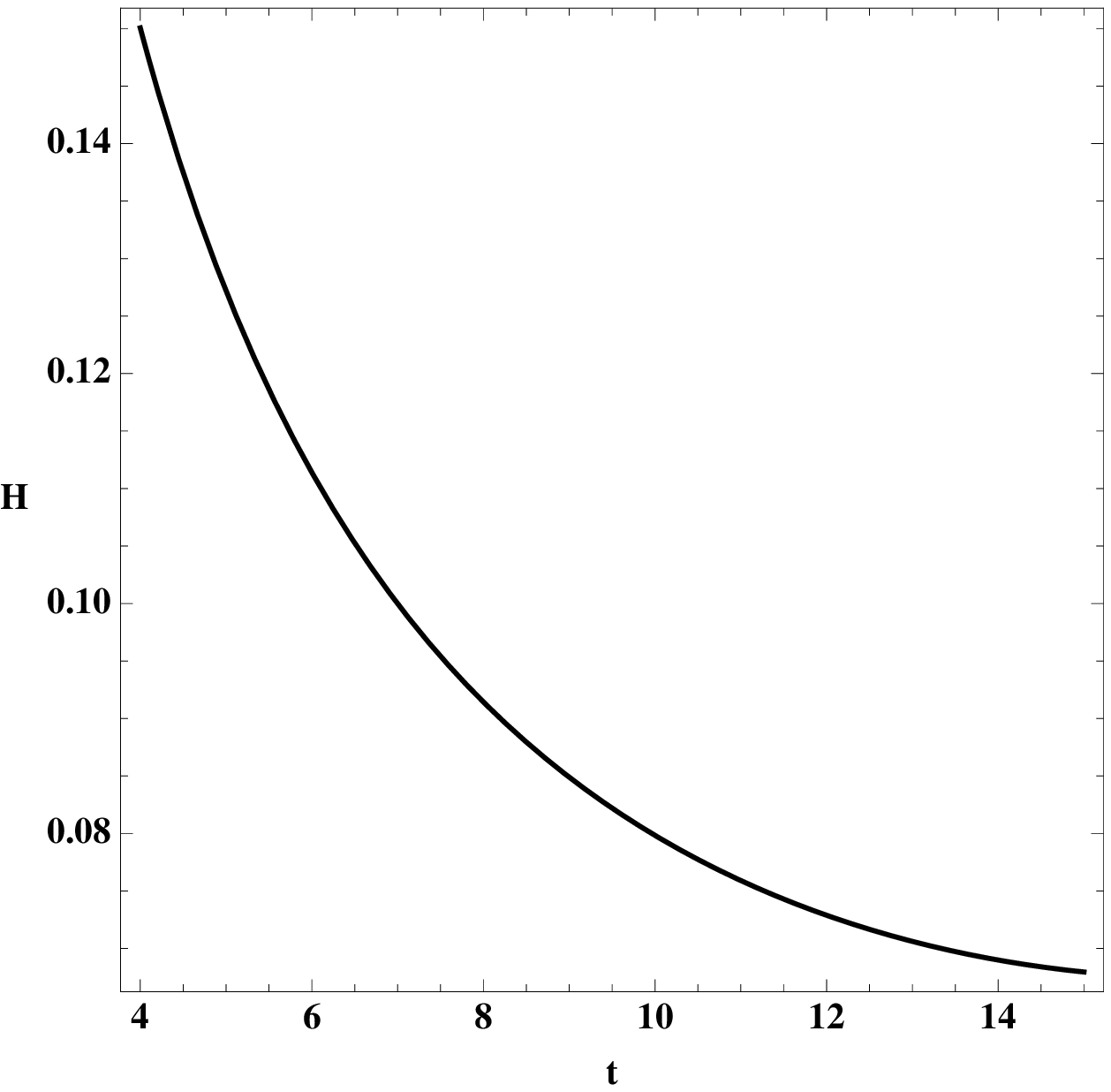}
\end{minipage} \hfill
\begin{minipage}[t]{0.2\linewidth}
\includegraphics[width=\linewidth]{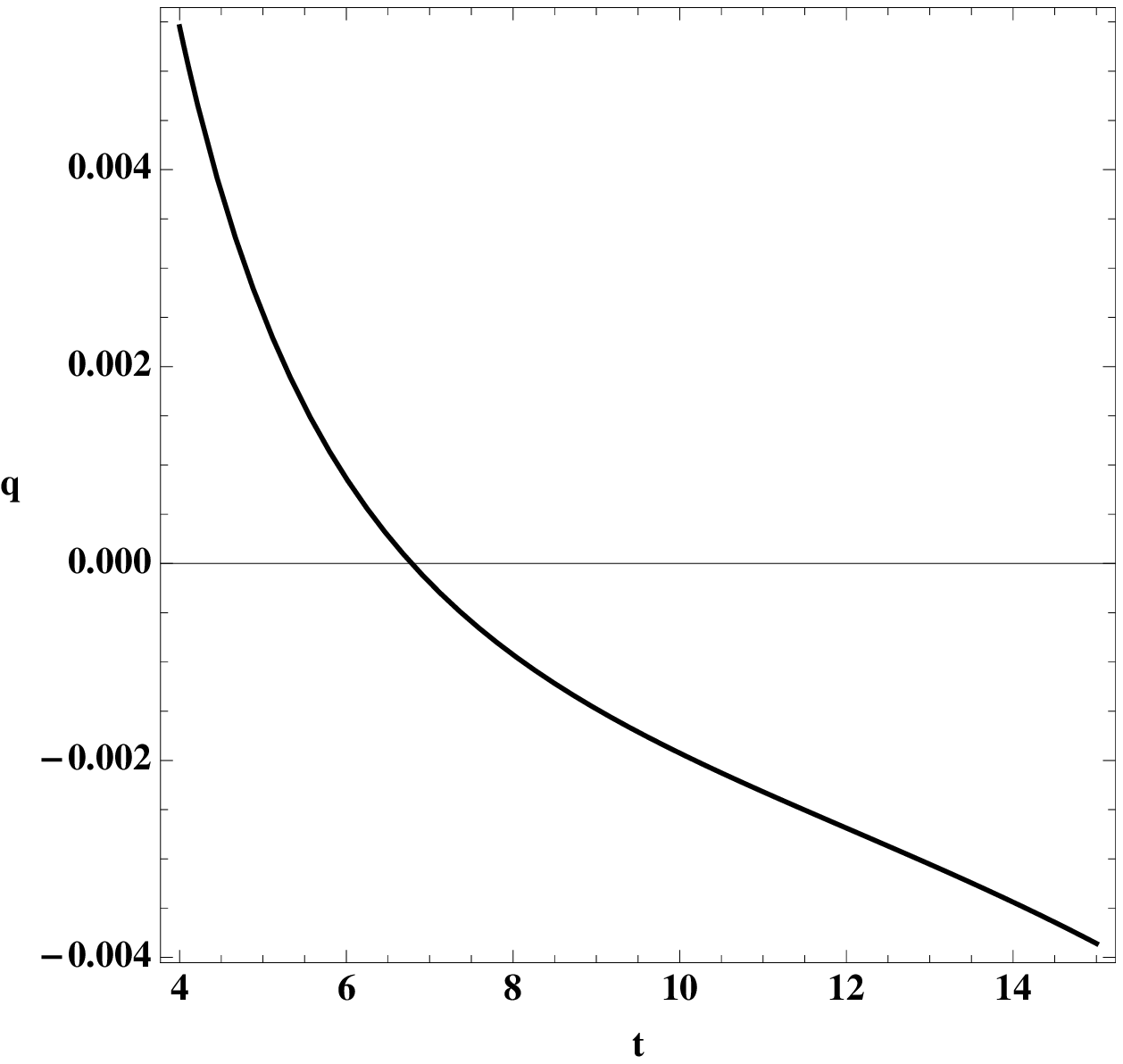}
\end{minipage} \hfill
\caption{The left panel represents the first root for $s$ displaying acceleration of the scale factor, the red line separating positive (left) and negative (right) values. The left center panel represents the first root for $p$, the red line separating positive (second and fourth quadrants) and negative values (first and third quadrants). The behaviour of the Hubble function $H$ and deceleration parameter $q$ for $\omega = 1$, $\lambda = - 1$ and $\alpha = 0$ are displayed in the right center and right panels.}
\end{figure}
\end{center}

\section{Conclusions}\label{conclusions}

In this work we have combined the idea of a scalar-tensor theory of the Brans-Dicke type and Rastall's proposal of a gravitational anomaly encoded in the violation of the conventional conservation law for the energy-momentum tensor. In doing so, we end up with two free parameters: the usual Brans-Dicke parameter
$\omega$ and Rastall's parameter $\lambda$, representing a degree of the non-conservation. 
\par
We have investigated the BDR theory in two contexts: spherically symmetric static solutions and cosmological regime. In the first case, we found that the only possible non-trivial analytical solution is a Robinson-Bertotti type solution.
The only possible solution in the BDR theory
that can represent a star is the usual Schwarzschild solution corresponding to the trivial configuration where the scalar field is constant. 
\par
For the cosmological case, we found power law solutions for the matter dominated phase, some of them representing an accelerating expansion, others, decelerating. We have shown that a decelerating/accelerating transition can be achieved in the matter dominated phase in the BDR theory. \newline
\vspace{0.2cm}
\newline
\noindent
{\bf Acknowledgements:} We thank FAPES (Brazil) and CNPq (Brazil) for partial financial support.


\begin{references}
\reference C. Brans and R.H. Dicke, Phys. Rev. {\bf 124}, 925(1961).
\reference A. Avilez and C. Skordis, {\it Cosmological constraints on
Brans-Dicke theory}, ArXiv:1303.4330.
\reference J.E. Lidsey, D. Wands and E.J. Copeland, Phys. Rep. {\bf 337},
343 (2000).
\reference L.E. Gurevich, A.M. Finkelstein and V.A. Ruban, Astrophys. Spac.
Sci. {\bf 22},
231(1973).
\reference A.B. Batista, J.C. Fabris, and R. de Sa Ribeiro, Gen. Rel. Grav.
{\bf 33}, 1237(2001).
\reference P. Rastall, Phys. Rev. {\bf D6}, 3357(1972); Can. J. Phys.
{\bf 54}, 66(1976).
\reference L.L. Smalley, Phys. Rev. {\bf D9}, 1635(1974).
\reference  B. Bertotti, Phys. Rev. {\bf 116}, 1331 (1959); I. Robinson, Bull.
Acad. Pol. Sci. {\bf 7}, 351 (1959).
\reference T.R.P. Caram\^es, M.H. Daouda, J.C. Fabris, A.M. Oliveira, O.F. Piattella and V. Strokov, Eur. Phys. J. {\bf C74}, 3145 (2014). 
\end{references}
\end{document}